# Isolated Ballistic Non-Abelian Interface Channel

Bivas Dutta[1], Vladimir Umansky[1], Mitali Banerjee[2] and Moty Heiblum[1,*]

[1]*Braun Center for Sub-Micron Research, Department of Condensed Matter Physics, Weizmann Institute of Science, Rehovot, Israel 76100*

[2]*Institute of Physics, Faculty of Basic Sciences, École Polytechnique Fédérale de Lausanne, Lausanne 1015, Switzerland*

* Corresponding author. Email: moty.heiblum@weizmann.ac.il

**Non-abelian anyons are prospective candidates for fault-tolerant topological quantum computation due to their long-range entanglement. Curiously these quasiparticles are charge-neutral, hence elusive to most conventional measurement techniques. A proposed host of such quasiparticles is the $\nu$=5/2 quantum Hall state. The gapless edge modes can provide the topological order of the state, which in turn identifies the chirality of the non-abelian mode. Since the $\nu$=5/2 state hosts a variety of edge modes (integer, fractional, neutral), a robust technique is needed to isolate the fractional channel while retaining its original non-abelian character. Moreover, a single non-abelian channel can be easily manipulated to interfere, thus revealing the state's immunity to decoherence. In this work, we exploit a novel approach to gap-out the integer modes of the $\nu$=5/2 state by interfacing the state with integer states, $\nu$=2 & $\nu$=3([1]). The electrical conductance of the isolated interface channel was 0.5e$^2$/h, as expected. More importantly, we find a thermal conductance of $0.5\kappa_0 T$ (with $\kappa_0$=π$^2$k$_B$$^2$/3h), confirming unambiguously the non-abelian nature of the $\nu$=1/2 interface channel and its Particle-Hole Pfaffian topological order. Our result opens new avenues to manipulate and test other exotic QHE states and braid, via interference, the isolated fractional channels.**


The fractional quantum Hall effect (FQHE) harbors fractionally charged quasiparticles localized in the 2D bulk surrounded by conducting chiral edge modes at the periphery([2]). Laughlin's states and their 'particle-hole conjugated' states are abelian([2]). In higher Landau levels, exotic non-abelian states are expected, each with a highly degenerate ground-state([3], [4]). A proposed state that hosts such exotic quasiparticles is the $\nu$=5/2 state, with gapless edge-modes: two integer modes ($\nu$=2), a fractional mode ($\nu$=1/2), and neutral Majorana modes (their number and chirality depend on the state's topological order). To capture the nature of the state, we recently employed thermal conductance $G_{th}$ measurement([5]), which is sensitive to all energy-carrying modes (charged and neutral alike). The measured $G_{th}$ is a vital topological invariant that dictates the state's topological order when all edge modes are fully thermal-equilibrated.



In the abelian regime, with fully equilibrated downstream (DS) and upstream (US) modes, the thermal conductance is given by $G_{th}=(n_d - n_u)\kappa_0 T$, with $\kappa_0 = \pi^2 k_B^2/3h$ the quantum of thermal conductance, $k_B$ the Boltzmann's constant, $h$ the Planck's constant, $T$ the temperature, and $n_d$ ($n_u$) is the number of downstream (upstream) edge modes. However, in the absence of thermal equilibration, one expects $G_{th}= (n_d + n_u)\kappa_0 T$ ([6]). In the non-abelian regime $G_{th}$ comes with fractional multiples of $\kappa_0 T$, originating from the fractional nature of the chiral central charge([7]). With this ambiguity in mind, we found experimentally $K_{5/2}=2.5\kappa_0$ ([5]), suggesting the Particle-Hole Pfaffian (PH-Pf) topological order of the state, disagreeing with the numerical calculations([8-23]). However, the nature of the *isolated* Majorana edge mode, tied with the $v=1/2$ charge mode, was never confirmed.

Here, we demonstrate the employment of a method we used to separate the fractional mode from its integer companions([1]). Our realization exploited an interface between two adjacent quantum Hall states([1]), with an *isolated channel* emerging at the interface. In particular, the interface between the $v=5/2$ state and the gate-defined $v=2$ or $v=3$, leads to a chiral or an anti-chiral $v=1/2$ interface modes. We found a thermal conductance coefficient of the isolated channel to be $K_{1/2}=0.5\kappa_0$ – indicating the non-abelian nature of the isolated channel, with its topological order agreeing with the PH-Pf([5]).

Molecular Beam Epitaxy (MBE) grown high quality GaAs-AlGaAs heterostructures with shallow-DX-centers doping([24]), allowing 'hysteresis free' gating and negligible heat conductance of the bulk at millikelvin temperatures([1]). This particular doping leads to higher disorder in comparison to the conventional superlattice (SPSL) doping technique([5], [24], [25]). Our experimental setup, with the 'heart' of the fabricated two-arm device is shown in Fig. 1A. The inner gates (yellow) and the outer ones (violet) tune the filling factor of each of the two bulks. The gates are separated from the sample by about a 25nm thick HfO$_2$ layer (for more details, see SM Sec. II). Gate voltage in the range -1.5V<$V_g$<+0.3V allows varying the electron density from pinch-off to 3x10$^{11}$cm$^{-2}$ (see Fig. S1); thus controlling the filling factor. The resulting interface modes split when leaving the small floating ohmic contact (20x2µm$^2$, shown in red), which connects the two arms. This contact serves as a heated reservoir. Large ohmic contacts (S, D, G, shown in cyan) are placed at the interfaces, while separate contacts, located at the physical edge of the mesa (not shown), probe the filling factor of each of the respective bulks.

Current $I_S$ is injected from S and dissipates power $\Delta P = \frac{1}{4}I_S^2 R_S$ in the floating contact ($R_S$ the two-terminal resistance of the interface). Figure 1B provides a schematic representation of the heat balance in the floating contact at the interface. Heat leaving the floating contact is composed of a phononic contribution $\Delta P_{ph}=\beta(T_m^5 - T_0^5)$ and an electronic contribution $\Delta P_e = 0.5K(T_m^2 - T_0^2)$, where $T_m$ is the temperature of the floating contact and $T_0$ the base temperature, with $\Delta P = \Delta P_{ph} + \Delta P_e$ at equilibrium.



For our small floating reservoir, the phononic contribution is negligible at $T_m$<20mK. The temperature $T_m$ is determined by measuring the 'low frequency' Johnson-Nyquist (J-N) thermal noise at the drain contact D, which is placed about 160$\mu m$ away along downstream in one arm (see SM Sec. III). The noise is filtered by an LC resonant circuit (resonance frequency 630 kHz and bandwidth 10-30kHz), amplified by a low-noise cold voltage pre-amplifier (placed on the 4.2K plate), followed by a room-temperature amplifier.

In general, our experimental strategy is to tune the inner regions to the 'tested' state $\nu_{in}$ (e.g., 5/2), and the outer regions to an integer state $\nu_{out}$ (e.g., 0, 1, 2, 3); leading to an 'interface filling' $\nu_{int} = \nu_{in} - \nu_{out}$ (with the nomenclature '$\nu_{in} - \nu_{out}$', used in all figures). The chirality of the interface charge mode reverses when the interfacing condition changes from $\nu_{in} > \nu_{out}$ to $\nu_{in} < \nu_{out}$. Having a single amplifier located downstream, the magnetic field was reversed between these two cases (see SM Sec. III).

We start with measurements of interface modes formed in the configurations '3-2', '3-1' and '3-0'. The two-probe source resistance $R_S$ exhibits quantized plateaus as a function of the gate voltage, $h/e^2$, $h/2e^2$, and $h/3e^2$ (~0.1% accuracy), with the expected thermal conductances $1\kappa_0 T$, $2\kappa_0 T$, and $3\kappa_0 T$, respectively(26, 27). In these chiral configurations, the downstream modes are equilibrated right from the start at the reservoir temperature $T_m$, which is determined by measuring the thermal J-N noise ($S_{th}$) at the drain (See SM Sec. III). The J-N noise and the deduced $T_m$ are plotted in Figs. 2A & 2B. We analyze the data in two ways: *i*. By fitting the linear-electronic contribution, $\Delta P$ versus $T_m^2$ for $T_m$<18mK, and find $K \cong 1\kappa_0$, $2\kappa_0$, and $3\kappa_0$, respectively (Fig. 2C); *ii*. By fitting the data for $T_m$<30mK (Fig. 2D), which reflects the electronic and phononic contributions. A similar quantization of the thermal conductance is found with $\beta \approx 5 \times 10^9 WK^{-5}$ (5).

Interfacing $\nu$=7/3, 5/2, and 8/3 with $\nu$=2 and $\nu$=3, leads to a source resistance shown in Fig. 3A. Well-quantized plateaus indicate full charge equilibration at the interfaces. We start with the $\nu$=7/3 and 8/3 abelian states. Interfacing these two states with $\nu$=2 results in an effective modes' filling of $\nu$=1/3 for '7/3-2' interface and $\nu$=2/3 mode for '8/3-2' interface (the latter is accompanied by an upstream neutral mode). The interface chirality is reversed when these states are interfaced with $\nu$=3: the $\nu$=2/3 charge mode accompanied by a neutral mode appears for '7/3-3' interface and a single $\nu$=1/3 charge mode appears for the '8/3-3' interface. The measured two-terminal thermal conductances in these four cases are shown in Figs. 3B & 3C. The results agree with good accuracy with the theoretical expectations(5).

The $\nu$=5/2 state may host different topological orders, with each having different modes structure and a different thermal conductance(28). Here, we consider the two competing candidates: Particle-Hole-Pfaffian (PH-Pf) and Anti-Pfaffian (A-Pf). For a discussion of other possible orders, see SM Sec. VI. Both orders support counterpropagating modes at the bare edge of the sample or, equivalently, the '5/2-0'



configuration (Figs. 4A-4B). Aside from two downstream integer charge modes and a $\nu=½$ charge mode, the PH-Pf order supports an upstream Majorana mode, while in comparison, the A-Pf order supports three upstream Majorana modes (two Majorana modes form a single bosonic neutral mode). With counterpropagating modes, an accurate determination of the thermal equilibration between all modes is imperative. With a full (none) thermal equilibration, one expects thermal conductance $2.5\kappa_0 T$ ($3.5\kappa_0 T$) for the PH-Pf and $1.5\kappa_0 T$ ($4.5\kappa_0 T$) for A-Pf.

Figures 4A-4F present different configurations: '5/2-0', '5/2-1', '5/2-2'. As shown in the figures, the expected two-terminal thermal conductance of the PH-Pf and A-Pf may overlap considering the range from full thermal equilibration to no equilibration. However, the interface '5/2-3' allows a clear distinction between the two topological orders (Fig. 4G-4H). Here, the interface channel for the A-Pf order supports two co-propagating modes, both leaving the hot reservoir at temperature $T_m$ (hence, equilibrated) with an expected thermal conductance $1.5\kappa_0 T$. By contrast, the PH-Pf order supports an interface channel with counterpropagating charge and a Majorana mode, leading to thermal conductance in the range $0.5\kappa_0 T - 1.5\kappa_0 T$.

In Figs. 4I-4L, we present the measurement results of $\Delta P$ vs $T_m^2$ for the four 5/2-interfaces: '5/2-0', '5/2-1', '5/2-2' and '5/2-3'. The measured thermal conductances are $(2.55\pm0.07)\kappa_0 T$, $(1.53\pm0.04)\kappa_0 T$, $(0.55\pm0.02)\kappa_0 T$, and $(0.53\pm0.02)\kappa_0 T$, respectively. These results, especially for the decisive-'5/2-3' interface, rules out the A-Pf order and point at the fully equilibrated PH-Pf order (Figs. 4A-4H). Above all, the observed ½-quanta thermal conductance of the fully equilibrated-isolated interface $\nu=½$ and Majorana channel, proves unambiguously its non-abelian nature.

The $\nu=5/2$ state has been attracting a wide spread attention due to growing expectations of realizing non-abelian anyons in condensed matter systems. But only proving the presence of non-abelian character of the state is not enough unless there is a robust sustainable method that allows isolating the non-abelian part from the trivial ones. Our work exploits a novel '*interfacing*' method (1), which allows measuring the thermal conductance of the emergent 'interface channel' at the junction between two adjacent QHE states; the 5/2 state and the integer states 2 and 3. The advantage of the interfacing method lies in the elimination of the integer modes and thus allowing a direct measurement of the isolated non-abelian interface channel. Moreover, this method allows a clear distinction between competing orders of the $\nu=5/2$ state. This configuration can be readily applied for interfering the singled-out non-abelian channel, thus allowing for the braiding operations - a necessary step in topological quantum processing.

**Acknowledgments**

We acknowledge D. Mross for a critical reading of the manuscript, followed by useful suggestions. B.D. acknowledges the support from Clore Foundation. M.H. acknowledges the continuous support of the Sub-Micron Center staff, the support of the European Research Council under the European Community's Seventh Framework Program (FP7/2007-2013)/ERC under grant agreement number 713351, the partial support of the Minerva foundation under grant number 713534.


**Author contributions**

B.D., M.H. designed the experiment, B.D. fabricated the devices and performed the measurements and did the analysis. B.D, M.B, and M.H participated in discussions and presenting the data. V. U. developed and grew the heterostructures supporting the 2DEG. All authors contributed to the manuscript writing.

**Competing interests**

The authors declare no competing interests.



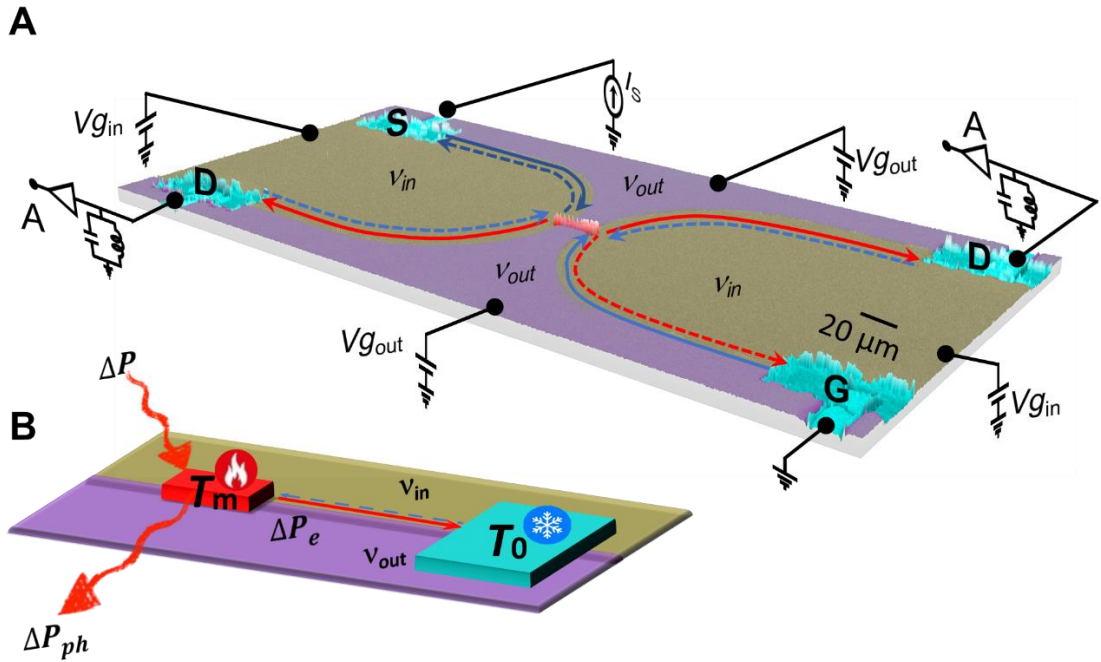

**Figure 1 | The experimental setup used to measure the interface thermal conductance. (A)** False colored SEM micrograph of the heart of the device. Two biased top gates, inner-gate ($Vg_{in}$, yellow) and outer-gate ($Vg_{out}$, violet), divide the 2DEG mesa into two gate-defined arms. A small floating ohmic contact (red, dimensions $20\times2\mu m^2$) connects the two arms. Large contacts (S, D, G, in cyan) probe the interface filling. Added ohmic contacts at the edge of MESA (not shown here) probe the filling of the respective sides. The floating contact is heated to $T_m$ by an injecting current $I_S$ from S. Its temperature is determined by measuring the low-frequency Johnson-Nyquist noise at 630kHz (LC bandwidth 10-30KHz), after amplification by a cooled pre-amplifier (at 4.2K) followed by a room-temperature amplifier. Arrows indicate the interface-modes at '5/2-3' interface as an example (see Fig. 4). **(B)** Schematic representation of heat balance in the floating contact, with the same color-codes as in **A**, showing only a zoomed-in part of the interface along downstream. An injected current dissipates power $\Delta P$, which is evacuated by two main channels: the phononic channel $\Delta P_{ph}$ and the electronic channel $\Delta P_e$. The cold reservoir contact G is at temperature $T_0$. Due to the small volume of the floating contact, the phononic contribution is negligible at low temperatures.
8

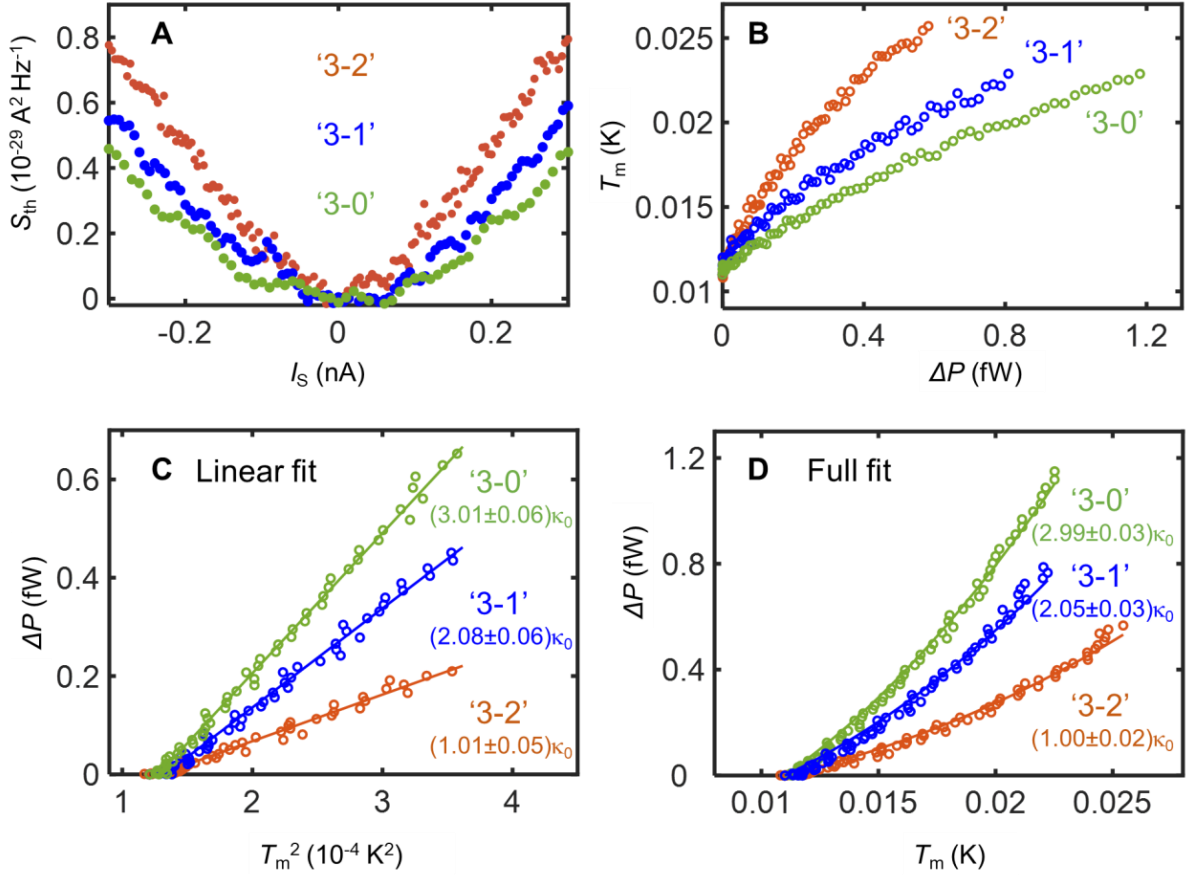

**Figure 2 | Noise, temperature, and dissipation at integer interfaces. (A)** Excess Johnson-Nyquist noise $S_{th}$ as function with heating current $I_S$ for the three 'integer interfaces': '3-2' (orange), '3-1' (blue), and '3-0' (green) at $T_0$=11mK. **(B)** Calculated temperature $T_m$ as a function of the dissipated power $\Delta P = 0.25 I_S^2 R_S$, with $R_S$ the 'interface mode' resistance, for the three different configurations (see SM Sec. V). **(C)** Dissipated power as a function of squared temperature in the range $T_m$<18mK. The thermal conductance is determined from the slope, via $\Delta P = \frac{1}{2} K (T_m^2 - T_0^2)$. **(D)** Dissipated power as a function of temperature in the range $T_m$<30mK. Here, phonon contribution is included, via $\Delta P = \frac{1}{2}\kappa(T_m^2 - T_0^2) + \beta(T_m^5 - T_0^5)$, with e-ph coupling constant β ≈ 5x10$^9$ WK$^{-5}$.



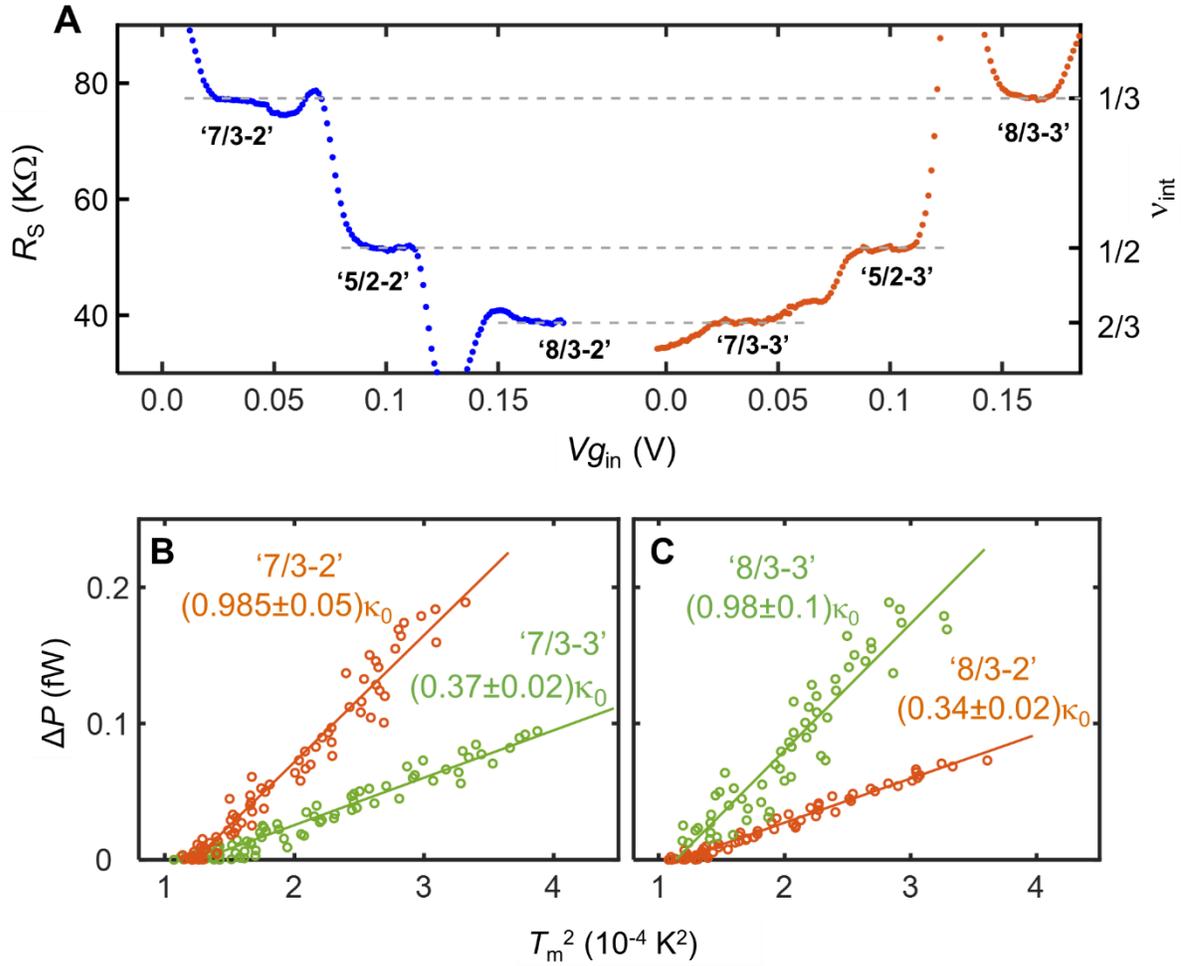

**Figure 3 | Interface resistance and thermal conductance of states in the 2nd LL. (A)** Two terminal source interface resistance of '7/3-2&3', '5/2-2&3', and '8/3-2&3'. Quantized plateaus indicate full charge equilibration at the interface. The peaks and dips are attributed to reentrant effects(29). **(B)** Determination of the thermal conductance for $T_m$<18mK, as was done in Fig. 2. The slope for '7/3-2' is close to unity, as expected for the 1/3 interface mode. The slope for '7/3-3' is $0.37\kappa_0$, and not *zero* - due to the well-known lack of equilibration of the resultant 2/3 mode(26). **(C)** Similar data as in **B**, at different interfacing conditions with $\nu$=8/3.



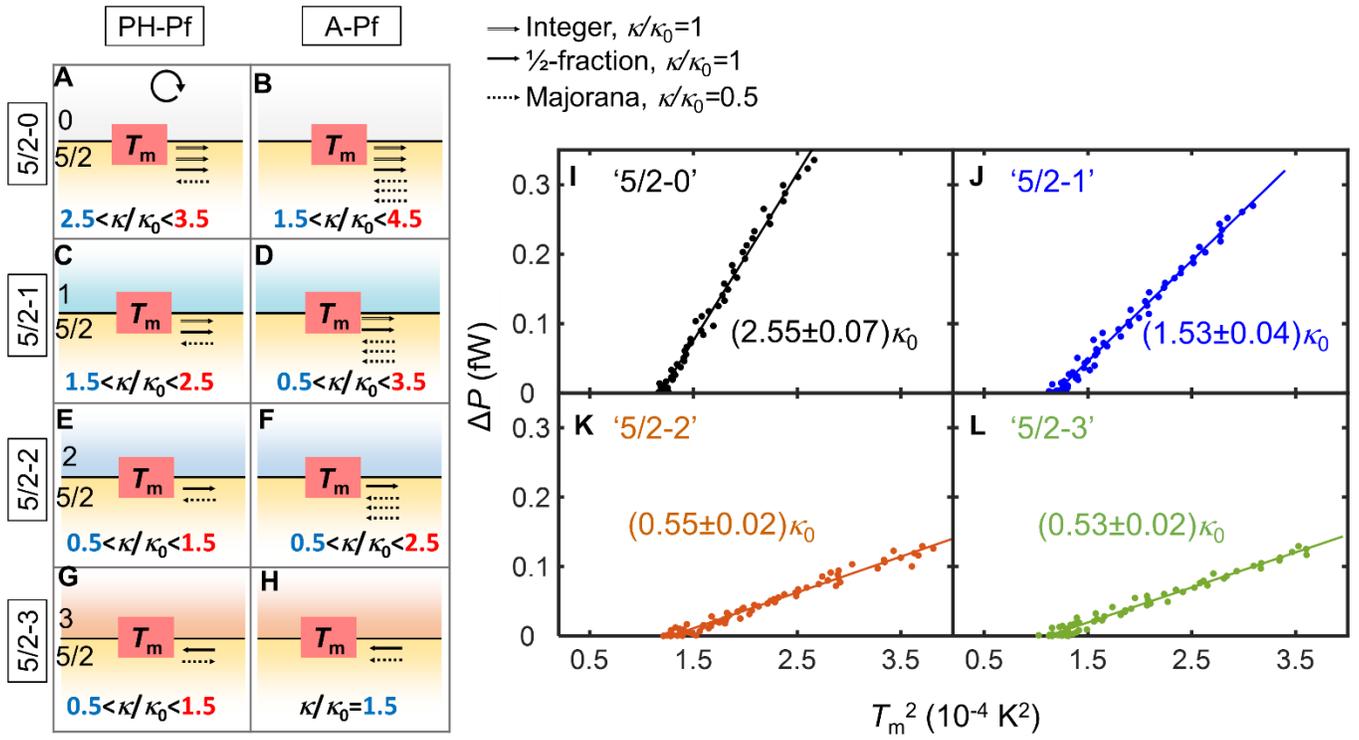

**Figure 4 | Thermal conductance of the '5/2-n' interfaces. (A-H)** Plotted are the interface modes of the interfaced PH-Pf and the A-Pf orders with different integers, and the expected thermal conductance in each case. Fully-equilibrated value in light-blue and un-equilibrated value in red. Notation of the arrows are indicated in the inset. Clockwise chirality is indicated by the circled arrow. **(I-L)** Plots of the dissipated power as a function of the squared temperature with linear fits as in Figs. 2 & 3, for all four 5/2-interfaces. The extracted thermal conductance values, in particular for '5/2-3', unambiguously exclude the A-Pf topological order (see **G-H**), and agrees (with small estimated errors) with the Ph-Pf topological order.



# Supplementary Materials: Isolated Ballistic Non-Abelian Interface Channel


Bivas Dutta[1], Vladimir Umansky[1], Mitali Banerjee[2] and Moty Heiblum[1,*]

[1]*Braun Center for Sub-Micron Research, Department of Condensed Matter Physics,*

*Weizmann Institute of Science, Rehovot, Israel 76100*

[2]*Institute of Physics, Faculty of Basic Sciences, École Polytechnique Fédérale de Lausanne,*

*Lausanne 1015 Switzerland*

\* Corresponding author. Email: moty.heiblum@weizmann.ac.il


## I. MBE-grown '*inverted*' heterostructure

The heterostructures used in these experiments were designed to achieve a compromise between the robustness of the 5/2 fractional state and the ability to operate it with surface gates. It is well established that a quantization of the 5/2 state is governed by the long-range spatial potential landscape formed by the ionized donors. Smoothing this landscape is customarily achieved by excessive doping, either in 'short period superlattice'(24), or in a low Al-mole-fraction in AlGaAs(5). These doping schemes provide a significant reduction in the long-range potential fluctuations via spatial correlations between ionized donors, exhibiting, at the same time, negligible lateral parasitic conductance at low temperatures. Unfortunately, weakly localized electrons in such doped layers both thwart the stable operation of the surface gates and completely inhibit applying positive gate bias, which is essential in the present experiment. In order to solve the problem, we use the so-called 'inverted' 2DEG structure, where an accurate quantization of the 5/2 state is enabled by excess δ-doping in a layer of $Al_xGa_{1-x}As$ (x=0.23) placed 60nm below the QW (see Fig. S1). The surface potential is compensated using a uniformly doped layer made of $Al_{0.37}Ga_{0.63}As$:Si, located far away from the QW, thus minimizing the impact of its long-range random potential fluctuations on the 2DEG. Such high Al mole-fraction doped layers are conventionally used between the gates and the 2DEG in most devices since the electrons freeze at T~100K in the DX-Si centers. The electron density in as-grown samples at 300mK is ~$2.4 \times 10^{11} cm^{-2}$ and the mobility is about $15 \times 10^6 cm^2$/Vs, even though the 2DEG wavefunction is asymmetrically shifted towards the bottom AlGaAs-GaAs interface - usually considered to be of lower quality than the top interface of the quantum well.



## II. Device fabrication

We start with a 250x800μm MESA, made by wet etching in $H_2O_2:H_3PO_4:H_2O=1:1:50$ solution for 2 mins resulting with etched depth of about 180 nm. In the next step, a few ohmic contacts are patterned by e-beam-lithography at the edge and in the bulk of the MESA, followed forming the ohmic contacts by metal deposition (with standard material ratio Au:Ge:Ni = 2:1:0.75), followed by annealing at 450°C for 2 mins. Some of the contacts are shown in Fig. 1a as S, G, D contacts. The whole sample is then coated with 25 nm of $HfO_2$ layer via ALD, acting as dielectric medium for the metallic-gate that follows. In the next step, the outer-gate ($Vg_{out}$) is patterned by e-beam-lithography and subsequently deposited a 20 nm Ti/Au thin metallic film, acting as the outer-gate. We then coat the sample with another 15 nm thick of $HfO_2$ layer, which separates the outer-gate from the following inner-gate. Next step consist of patterning of inner-gate ($Vg_{in}$) and subsequent metallization by 20 nm Ti/Au thin film. The ohmic contacts (such as S, G, D) in the bulk of the MESA were designed such that they sit exactly at the interface of the two gates. In the final step, the interface-ohmic contacts (like, S, D, G contacts) are connected to the bonding-pads by thick gold lines, passing over the $HfO_2$ covered outer-gate, such that they do not short to the gates.

## III. Experimental setup

The heart of the device is shown in Fig. 1a. Gate voltages ($Vg_{in}$ and $Vg_{out}$) are applied to divide the mesa into two gate-defined arms. The small-floating ohmic contact with area 20x2 μm$^2$ act as the floating hot-reservoir connecting the two arms. An injected DC current $I_s$ from the interface-contact 'S' heats up the floating contact to temperature $T_m$. The voltage amplifiers are connected to the drain contact 'D' at the interface. Thermal voltage fluctuations ($S_v$) of the hot-floating contact is measured by the amplifiers after being filtered by the LC resonant circuit. The voltage fluctuations are converted into current fluctuations as $S_{th} = S_v G_{int}^2$, with $G_{int}$ the conductance of the interface mode.

In the following we give an example, which describes the experimental process for the measurements at the 5/2-interfaces. First, the magnetic field is tuned to a value B≈(-/+) 4T (chirality clockwise/anti-clockwise). At this magnetic field, with the bare density (n~2.4×10$^{11}$cm$^{-2}$) the whole sample remains at a filling ν>2, just outside the ff=2 plateau. Now, we create 5/2-interfaces by tuning the gates. For making '5/2-n' interface, where, n=2,1,0 (at B=-4T), we apply positive gate voltage $Vg_{in}$ to populate the inner-gated region to $\nu_{in}$=5/2, while apply a negative gate voltage $Vg_{out}$ on the outer-gate to deplete the outer-gated region to $\nu_{out}$=2, 1, or fully pinch to $\nu_{out}$=0. Similarly, for making '5/2-3' interface (at B=+4T; with a fixed amplifier position we needed to reverse B to +4T, to have the amplifier on the downstream side), we apply



certain positive gate voltages on both gates to populate the inner-gated regions to $\nu_{in}$=5/2, and the outer-gated region to $\nu_{out}$=3, respectively. Secondly, to characterize the 'interface-mode', we measure the 2-terminal resistance at the interface ohmic contacts (e.g., at 'S'). Nicely quantized interface-resistance ensures proper charge equilibration. Next, we measure the branching of the impinging current at the floating Ohmic contact. We assure symmetric partitions of the current between the two arms, with less than 2% reflections. After this, we proceed for the thermal conductance measurements at the interface.

## IV. Determination of amplifier Gain and $T_0$

The gain of the amplifier is a crucial parameter to determine the temperature $T_m$ of the floating-contact. We take the advantage of the temperature dependent Johnson-Nyquist voltage noise of the well-known Hall-resistance $R_S$ of the interface modes to calibrate the gain the of amplifier. The equilibrium voltage noise of the $R_S$ (well-known with good accuracy) is measured at the central frequency ($f_0$~630KHz) of the LC resonant circuit, with a $\Delta f$ bandwidth opened at the spectrum analyzer, for different cryostat (bath) temperatures $T_{bath}$ measured by well calibrated commercial thermometers, $S_V = 4k_B R_S T_{bath} G_h^2 G_c^2 \Delta f + S_{const}$, where $k_B$, the Boltzmann constant; $G_h$=400, the known gain of the room-temperature amplifier; and $S_{const}$ is the temperature independent voltage noise of the amplifier. A plot of the normalized voltage noise, $S_{norm} = S_V/(4k_B R_S G_h \Delta f)$, at the '2-1' interface, with different bath temperatures is shown in Fig. S2. The slope of the linear fit of the data gives the gain of the cold amplifier, $G_c = \text{sqrt}(\Delta S_{norm}/\Delta T)$. The obtained gain at this interface is $G_c$~7.7, which is the effective gain of the amplifier depending on the particular bandwidth of the LC circuit. For different interface modes the bandwidth (depends on $R_S$) of the circuit changes and so the effective gain of the amplifier. The ratio of the areas of resonance curves for two interfaces is directly proportional to the ratio of the squared gains in the two cases. The area under the LC resonance curves for two interfaces (with two different $R_s$) were compared to extract the gain at different interface fillings.

After knowing the gain of the amplifier, the temperature $T_0$ is inferred from the above calibration curve, by measuring the thermal-background noise of the interface resistances.

## V. Determination of $T_m$

We extract the temperature $T_m$ of the small-floating contact by measuring the excess thermal fluctuations from the floating contact carried out by the interface edge modes. Using the general expression for the



excess current noise $S_{th}$, carried out by $n_1$ and $n_2$ number edge channels in the two arms, we get the temperature of the floating contact $T_m$ given by,

$$T_m = T_0 + S_{th}/2k_B G^*,$$

Where, $G^* = G_{int}\frac{n_2(n-n_2)}{n}$, with $n = n_1 + n_2$, and $G_{int}$ is the conductance of the interface mode. In our case $n_1 = n_2$, therefore, $G^* = ½ G_{int}$.

## VI.   5/2-interface thermal conductance for other possible topological orders

In the main text we have considered PH-Pf and A-Pf topological orders of $v$=5/2 state. The Pf order (Fig. S3), which supports a DS Majorana mode, is also numerically favorable. But the observation of US noise makes it experimentally unlikely. Still, at the interface of the Pf-order with $v$=2(Fig. S3) one expects a definite value of thermal conductance K$T$=1.5$\kappa_0 T$, which doesn't match the observed value in our experiment, K$T \cong 0.5\kappa_0 T$ (see Fig. 4K). Therefore, we can exclude the Pf order. Similarly all other a-priori possible topological orders can be excluded either by the interface with $v$=2 or with $v$=3 (28).

## VII.   Hall measurements near $v$=5/2 state

Fig. S4 shows the 2-probe Hall resistance $R_{xy}$ with magnetic field B near the $v$=5/2 state, measured at the physical edge of sample (with $v$=0 outside), showing a well quantized plateau.

## VIII.   5/2-interface thermal conductance with small variation of filling on the plateau and the base temperature $T_0$

The Fig. S5-S6 shows the data for the thermal conductance at the interface '5/2-3' with different filling factors $v_{in}$ around $v$=5/2 and at different base temperature $T_0$. The measured values found to be more or less constant with the changes of filling factors and temperatures, with $\kappa \approx 0.5\kappa_0$.

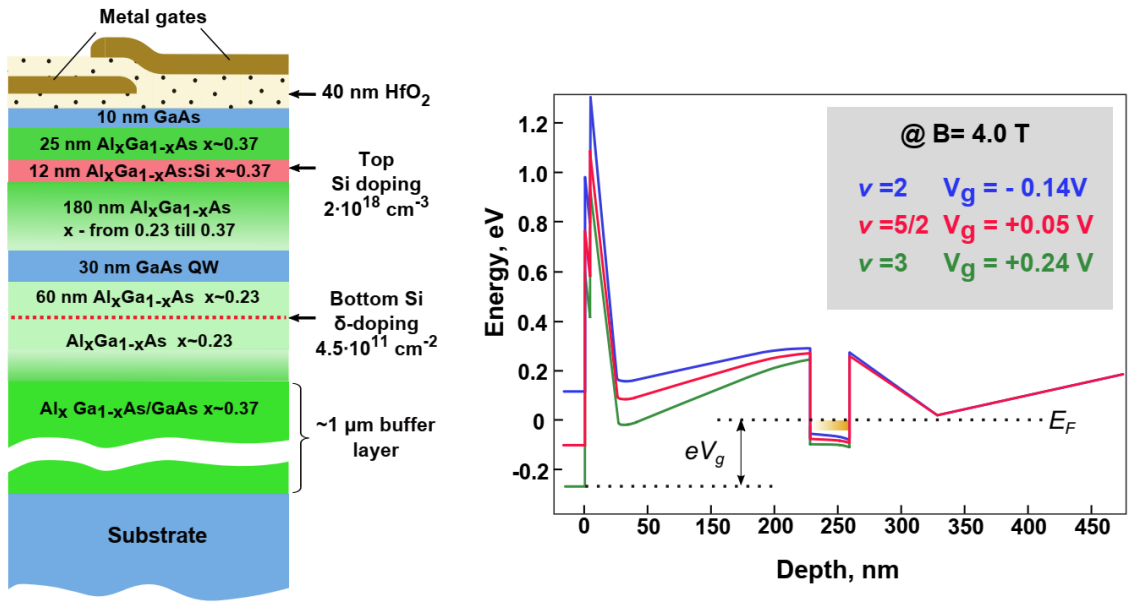

**Figure S1| MBE growth structure of the heterostructure. left,** Cross-section of the MBE growth (not in scale). **right,** Conduction band at three different gate voltages corresponding to three different filling factors $\nu$=2, 5/2, and 3.

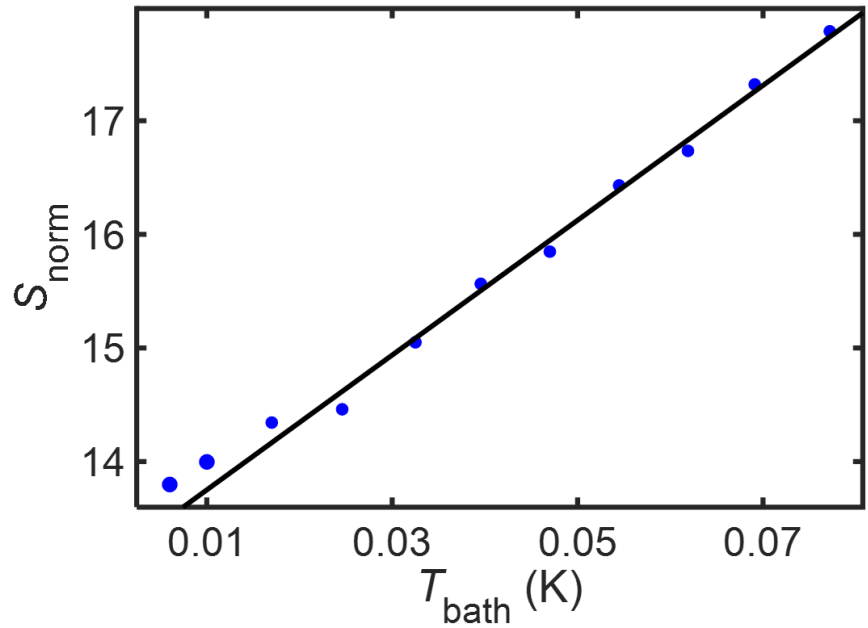

**Figure S2| Calibration of the amplifier.** Temperature-dependent Johnson-Nyquist noise of the interface resistance $R_S$ allows calibration of the amplifier. The normalized noise at the '2-1' interface is plotted with the bath temperature. The slope of the linear fit gives the gain of the amplifier for this interface.

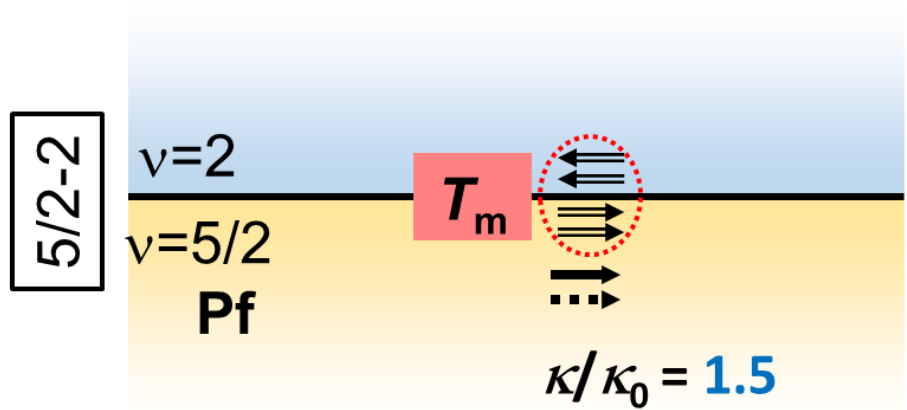

**Figure S3| expected interface thermal conductance for '5/2-2' interface with the Pf topological order.** At the interface of $v$=5/2-Pf order with nu=2, two integer modes are gapped out, leaving at the interface $v$=1/2 and a Majorana mode both in the DS, hence, thermally equilibrated at $T_m$. The expected thermal conductance is $1.5\kappa_0 T$, in disagreement with the experimentally observed value(Fig. 4k).

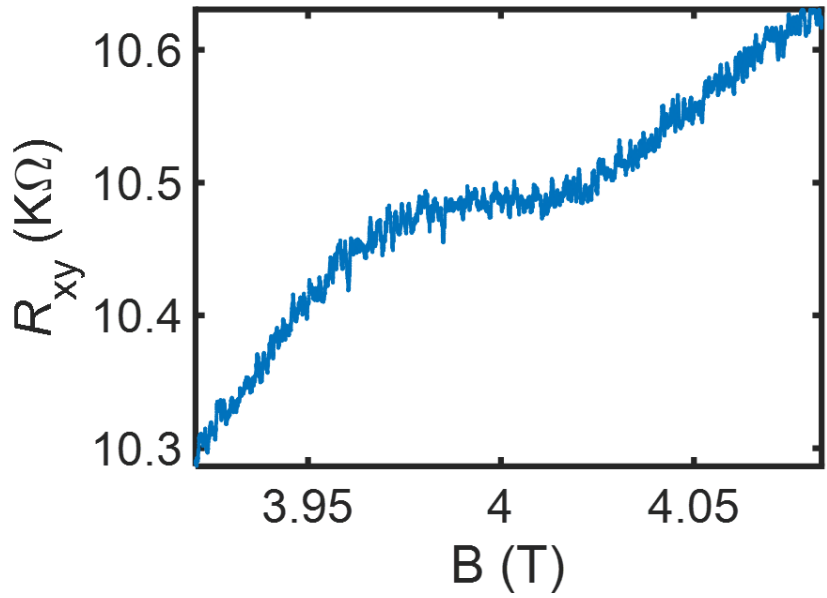

**Figure S4| Hall measurements near $v$=5/2.** 2-probe Hall resistance $R_{xy}$ with magnetic field B near the $v$=5/2 state, measured at the physical edge of sample. A well-quantized plateau with width of 500-600 G is observed, similar to previous samples(5). Note that, the 2-probe measurements include an extra contact resistance.

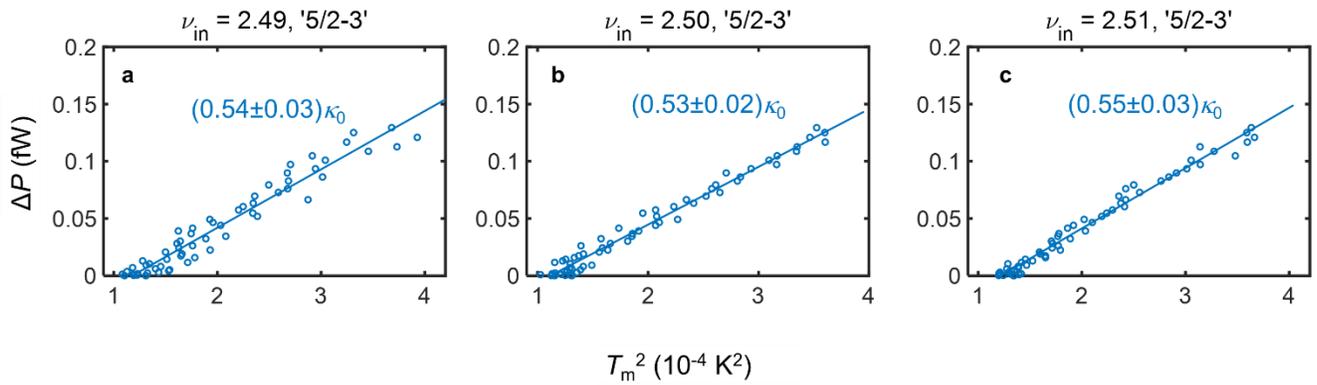

**Figure S5| '5/2-3' interface thermal conductance with different positions on the 5/2 plateau.** The ΔP vs $T_m^2$ data for the '5/2-3' interfaces are plotted with a variation of the inner-gate filling $\nu_{in}$ on the 5/2 plateau. The extracted thermal conductances for these three positions are found to be ~$0.5\kappa_0$ (within the estimated error-bar).

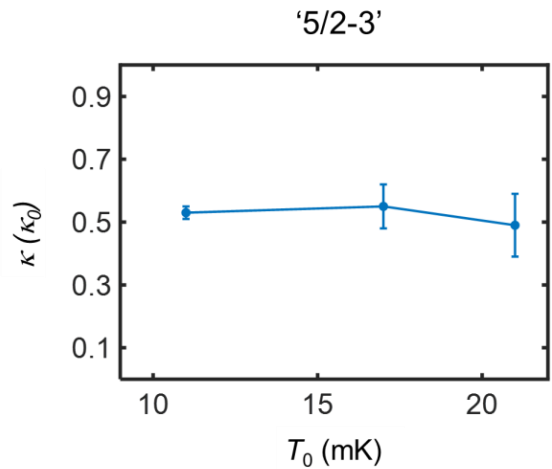

**Figure S6| Temperature dependence of the '5/2-3' interface Thermal conductance.** The thermal conductance of '5/2-3' interface is plotted for three different base temperature $T_0$ =11, 17 and 21mK, and found to be almost constant within the estimated error-bars with $\kappa \approx 0.5\kappa_0$.